\input harvmac
%\draft
\noblackbox
%-------------------------
% This paper uses harvmac
%-------------------------
\overfullrule=0pt
\def\Title#1#2{\rightline{#1}\ifx\answ\bigans\nopagenumbers\pageno0\vskip1in
\else\pageno1\vskip.8in\fi \centerline{\titlefont #2}\vskip .5in}

%
%-------------------
%  definitions
%
%
\def\s{\sigma}

\def\dE{ \Delta E }
\def\e{\epsilon}
\def\apm{}

\def\[{\left [}
\def\]{\right ]}
\def\({\left (}
\def\){\right )}
\def\dfourk{ {d^4 k \over (2 \pi )^4} }
\def\p{\partial}

\font\cmss=cmss10 \font\cmsss=cmss10 at 7pt
\def\IZ{\relax\ifmmode\mathchoice
   {\hbox{\cmss Z\kern-.4em Z}}{\hbox{\cmss Z\kern-.4em Z}}
   {\lower.9pt\hbox{\cmsss Z\kern-.4em Z}}
   {\lower1.2pt\hbox{\cmsss Z\kern-.4em Z}}\else{\cmss Z\kern-.4emZ}\fi}

%%-------------------
% references
%

\lref\arl{H. Araki and E. Lieb, Comm. Math. Phys. {\bf 18} (1970) 160.}
\lref\dasmathurtwo{S. Das and S. Mathur,
``{\it Interactions Involving D-branes}'',
hep-th/9607149.}

\lref\gk{ S. Gubser and I. Klebanov, 
``{\it Emission of Charged
Particles
from Four- and Five-dimensional Black Holes}'', 
hep-th/9608108.}

\lref\page{ D. Page, Phys. Rev. {\bf D 13} (1976) 198; Phys. Rev. {\bf
D14}
(1976), 3260.}

\lref\unruh{W. Unruh, 
Phys. Rev. {\bf D14} (1976) 3251.}
\lref\cj{ M. Cvetic and D. Youm, hep-th/9512127.}

\lref\tseytlin{A. Tseytlin, hep-th/9601119.}
\lref\tata{A. Dhar, G. Mandal and S. Wadia, ``{\it
Absorption vs Decay of Black Holes in String Theory and 
T-symmetry}'', hep-th/9605234.}
\lref\gm{G. Horowitz and D. Marolf, hep-th/9605224.}
\lref\mss{J. Maldacena and L. Susskind, ``{\it D-branes and Fat 
Black Holes }'',  hep-th/9604042.}
\lref\bl{V. Balasubramanian and F. Larsen, hep-th/9604189.}
\lref\jmn{J. Maldacena,``{\it Statistical Entropy of Near Extremal
Fivebranes}'', hep-th/9605016.}
\lref\hlm{G. Horowitz, D. Lowe and J. Maldacena, hep-th/9603195.}
\lref\ms{J. Maldacena and A. Strominger, hep-th/9603060.}
\lref\bd{See N. D. Birrel and P.C. Davies,``{\it Quantum Fields in
Curved
Space}'', Cambridge University Press 1982.}
\lref\hms{G. Horowitz, J. Maldacena and A. Strominger,
``{\it Nonextremal Black Hole Microstates and U-duality}'',
 hep-th/9603109.}
\lref\dbr{J. Polchinski, S. Chaudhuri, and C. Johnson, hep-th/9602052.}
\lref\jp{J. Polchinski, hep-th/9510017.}
\lref\dm{S. Das and S. Mathur, ``{\it
Comparing Decay Rates for Black Holes and D-branes}'',hep-th/9606185.}
\lref\ghas{G. Horowitz and A. Strominger,``{\it Counting States of
Near-extremal Black Holes }'',  hep-th/9602051.}
\lref\ascv{A. Strominger and C. Vafa, Phys. Lett. {\bf B379} (1996)
99,
 hep-th/9601029.}
\lref\hrva{P. Horava, Phys. Lett. {\bf B231} (1989) 251.}
\lref\cakl{C. Callan and I. Klebanov, hep-th/9511173.}
\lref\prskll{J. Preskill, P. Schwarz, A. Shapere, S. Trivedi and
F. Wilczek, Mod. Phys. Lett. {\bf A6} (1991) 2353. }
\lref\bhole{G. Horowitz and A. Strominger,
Nucl. Phys. {\bf B360} (1991) 197.}
\lref\bekb{J. Bekenstein, Phys. Rev {\bf D12} (1975) 3077.}
\lref\hawkirr{S. Hawking, Phys. Rev {\bf D13} (1976) 191.}
\lref\stas{A.~Strominger and S.~Trivedi,  Phys.~Rev. {\bf D48}
 (1993) 5778.}
\lref\bek{J. Bekenstein, Lett. Nuov. Cimento {\bf 4} (1972) 737,
Phys. Rev. {\bf D7} (1973) 2333, Phys. Rev. {\bf D9} (1974) 3292.}
\lref\hawkb{S. Hawking, Nature {\bf 248} (1974) 30, Comm. Math. Phys.
{\bf 43} (1975) 199.}
\lref\cm{C. Callan and J. Maldacena, Nucl. Phys. {\bf B 475 } (1996)
645, hep-th/9602043.}
\lref\hpc{S. Hawking, private communication.}
\lref\bdpss{T. Banks, M. Douglas, J. Polchinski, 
S. Shenker and A. Strominger, in progress.}
\lref\ast{A. Strominger, {\it Statistical Hair on Black Holes''},
hep-th/9606016.  }
\lref\ka{ B. Kol and A. Rajaraman, ``{\it
Fixed Scalars and Suppression of Hawking Evaporation}'',  hep-th/9608126. }

\lref\thooft{G. 't Hooft, 
``{\it The Scattering Matrix Approach for the Quantum Black Hole, 
an Overview}'', gr-qc/9607022.}

%-------------------
% title page
%-------------------
%
\Title{\vbox{\baselineskip12pt
\hbox{hep-th/9609026}\hbox{RU-96-78}}}
{\vbox{
\centerline {Black Hole Greybody Factors } 
\centerline{and D-Brane Spectroscopy}  }}
\centerline{Juan Maldacena}
\vskip.1in
\centerline{\it Department of Physics and Astronomy, Rutgers University,
Piscataway, NJ 08855}
\vskip.1in
\centerline{\it and}
\centerline{Andrew Strominger}
\vskip.1in
\centerline{\it Department of Physics, University 
of California,Santa Barbara, CA 93106}
\vskip .5in
\centerline{\bf Abstract}
Black holes do not Hawking radiate strictly blackbody 
radiation due to well-known frequency-dependent
greybody factors. These factors arise from frequency-dependent 
potential barriers outside the horizon which filter the initially  
blackbody spectrum emanating from the horizon. 
D-brane bound states, in a thermally excited state 
corresponding to near-extremal black holes, 
also do not emit blackbody radiation: The bound state 
radiation spectrum encodes the
energy spectrum of its excitations.
We study a near-extremal five-dimensional black hole.  We show that, 
in a wide variety of circumstances including both neutral 
and charged emission,
 the effect of the 
greybody filter is to transform the blackbody radiation 
spectrum precisely into the bound state 
radiation spectrum. Implications of 
this result for 
the information puzzle in the context of 
near-extremal black hole dynamics are discussed. 
 \Date{}

%
%----------------------
% Body of Paper
%----------------------
\newsec{Introduction}
In \ascv\ the Bekenstein-Hawking entropy formula was derived 
for certain five-dimensional 
extremal black holes in string theory 
by counting the asymptotic 
degeneracy of BPS-saturated D-brane bound states. This derivation 
required an extrapolation from the small black hole region,
where D-brane perturbation theory is good and the Schwarzchild 
radius is smaller than the string length, to the large black hole 
region where the low-energy semiclassical approximation and 
the Bekenstein-Hawking formula are valid.  The extrapolation was 
justified by the 
special topological character of BPS states, which implies that their 
degeneracies should not change under smooth variations of couplings.  
It was stated in \ascv\ that the use of D-brane perturbation theory 
to study large black holes was likely limited to such supersymmetric 
counting problems, and could not be extended to study 
dynamics of non-BPS excited states. 

However, this view proved to be too 
conservative: In \ghas\ the entropy of near-extremal states of large 
black holes
was found,
in the ``dilute gas" region\foot{Outside this region interactions between 
left and right moving oscillations cannot be neglected and the string 
calculations are difficult \cm\ \hms.} (defined in section 2),  to be 
completely accounted for by low-lying 
non-BPS oscillations of an effective string. We shall see that 
this effective string, which arises
% as the intersection 
%region 
in the description 
of bound D-branes \ascv , provides a very robust 
picture of extremal black hole dynamics. 
The entropy counting in \ghas\  
worked because the oscillations are highly diluted 
in the dilute gas region 
and potentially strong interactions between 
them are accordingly suppressed.  Decay of these excited 
states (i.e. Hawking radiation) occurs as oscillations 
dissipate into radiation \cm , and it was further noted \cm\ 
that the rate
had the roughly the right features. 
However,  in the large black hole region  computation of 
the string radiation rate 
appears to be a strong coupling problem. Hence it was stated in \ghas\
that string techniques were unlikely to give a precise calculation of 
the decay rate. 

However, this view also proved to be too conservative. The leading order decay 
rate of the thermally excited string 
into a single species of 
neutral S-wave scalars of frequency $\omega$ is given by 
\eqn\dhaw{
\Gamma_{D} = g_{eff}\omega 
\rho({ \omega \over 2 T_L })\rho({ \omega \over 2 T_R })\dfourk .}
$g_{eff}$ is a (charge-dependent but frequency-independent) 
effective coupling of left and
right moving oscillations 
of energies $\omega/2$ to
an outgoing scalar of energy $\omega$. 
$T_L$ and $T_R$ are the temperatures of left and
right moving oscillations, and are related to the overall
temperature $T_H$ by 
\eqn\tlr{{1\over T_R} +{1\over T_L}={2 \over T_H}.} 
The thermal 
factor $\rho(\omega/T)$ 
is
\eqn\rdef{\rho({\omega\over T})\equiv { 1 \over e^{ { \omega \over T } }  -1
}.}
These thermal factors arise in  \dhaw\ from the left and right moving 
oscillation densities. The black hole decay rate on the other hand is given
by the Hawking formula \hawkb\
\eqn\hgf{\Gamma_H=\sigma_{abs} (\omega) \rho({\omega \over T_H})
\dfourk .}
where $\sigma_{abs} (\omega)$ is the greybody factor, which 
equals the  classical absorption cross section. 
In the limit $T_R\ll T_L$ these equations simplify dramatically,
and both depend on the frequency as $\rho({\omega / T_H})$. 
It was shown in \cm\ \tata\  that, in this limit, 
 both $\Gamma_{D}$ and $\Gamma_H$ are 
proportional to the area and, in a surprising paper by Das and Mathur
\dm \dasmathurtwo , that the numerical coefficient also matches.
Note that this result is confined to the  near extremal 
region, in which the wavelength of the outgoing radiation is 
much larger
than the Schwarzchild radius. 

In this paper we consider the highly non-trivial comparison in which 
the restriction $T_R\ll T_L$ is 
dropped, while remaining in the dilute gas and near extremal 
regions. After a lengthy calculation 
we find that the semiclassical greybody factors are  
\eqn\grb{\sigma_{abs} (\omega)={g_{eff}\omega 
\rho({ \omega \over 2 T_L })\rho({ \omega \over 2 T_R }) \over
\rho({\omega \over T_H})} ,}
implying $\Gamma_D=\Gamma_H$ and exact agreement between the string 
and semiclassical calculations!.     

Let us summarize this. The black hole emits blackbody radiation from
the horizon. Potential barriers outside the horizon
act as a frequency-dependent 
filter, reflecting some of the radiation back into the black hole and
transmitting some to infinity. The filtering acts in 
just such a way  that the black hole spectroscopy mimics the excitation  
spectrum of the string. Hence  to the observer at infinity the
black hole, masquerading in its  greybody cloak, 
looks like the  string, for  energies small compared to the 
inverse Schwarzchild radius of the black 
hole.

In the past, greybody factors have been largely regarded as 
annoying factors which mar the otherwise perfectly thermal 
blackbody radiation. Now we see that they have an important place in
the order of things, and transmit a carefully inscribed message 
on the quantum structure of black holes.
We also see that in order to compare the string and black hole pictures,
we must take into account proccesses which occur well $outside$ the horizon 
of the black hole solution.
This is surprising in that D-brane 
bound states comprising the string are conventionally viewed as  
confined to a very small region. 

We further consider the case of charge emission\foot{The emission rate
in the limit $T_R \ll T_L$ was recently derived in \gk.}. The formulae
generalize the above with the appearance of an extra charge parameter.
It turns out that under some circumstances charge emission 
dominates neutral emission for a near-extremal black 
hole. Again we find exact
agreement between string and semiclassical results 
everywhere in the dilute gas region.  

The reason for this 
precise agreement remains mysterious. As shall be explained in Section 3,  
one calculation is 
an expansion in the size of the black hole, while the other is 
an expansion in the inverse size. A priori both were expected 
to get corrections and there 
was no obvious reason that they should agree. 
The agreement 
strongly suggests that 
there is much yet to be learned about these fascinating 
objects. 
Perhaps there is a 
supersymmetric 
nonrenormalization theorem 
protecting the interactions between BPS states from corrections, or
they are suppressed by our
 restrictions to low energies and or the dilute gas region.
 We
see no reason to expect the agreement to persist outside the near-extremal 
region when wavelengths are of order the Schwarzchild radius - 
but there could be more surprises!

In conclusion the string picture of black hole dynamics is apparently 
far more 
robust than originally envisioned in \ascv, at least 
when restricted to low excitation energies 
in the dilute gas region. 
The string 
decay rates, extrapolated to the large black hole region,
agree precisely with the semiclassical Hawking decay rates in 
a wide variety of circumstances. However, the string method
not only supplies the decay $rates$, but it also gives a set of unitary 
$amplitudes$ underlying the rates. We find it tempting to conclude that 
these extrapolated amplitudes are also correct. It is hard to imagine a 
mechanism which corrects the amplitudes, but somehow conspires to 
leave the rates unchanged.  

This robust nature of the string picture 
is very significant because it allows us to directly 
confront the black hole information puzzle, which is of course a
primary goal of these investigations. According to Hawking  
information is lost as a large excited black hole decays to extremality. 
On the other hand the string analysis - extrapolated to the 
large black hole region -  gives a manifestly unitary
answer. We will not reconcile these points of view but we will  make 
some hopefully relevant observations along the way.

In section 2 we review the classical black hole solution. 
In section 3 we discuss the semiclassical limit
and expansion parameters. In section 4 
we compute and compare the emission rates for neutral scalars 
using the Hawking and string methods. Section 5 
considers the charged case. Comparisons of absorption rates are made
in section 6. Section 7 discusses the rate of charge loss of a 
black hole and contains comments on measuring the quantum microstate 
by scattering experiments.

\newsec{The Classical Solution} 

In this section we collect some known properties of the 
classical five-dimensional black hole solutions and their 
D-brane descriptions which will be needed in the following. 
Except where otherwise noted, we adopt 
the notation of \hms\ including $\alpha^\prime =1$, 
so that all dimensionful quantities are measured in string units.
The low-energy action for ten-dimensional type IIB string theory
contains the terms, 
\eqn\fds{{1\over 16 \pi G_{10}}
\int d^{10}x \sqrt{- g} \[ e^{-2\phi}\(R+4(\nabla \phi )^2\)
-{1 \over 12} H^2 \]}
in the ten-dimensional string frame.
$H$ denotes the RR three form field
strength, and $\phi$ is the dilaton. The NS three form, self-dual five
form, and second scalar are set to zero.
We will let $g$ denote the
ten-dimensional string coupling and define the zero mode of
$\phi$ so that
$\phi$ vanishes asymptotically. The ten-dimensional Newton's constant is then
$G_{10}=8 \pi^6 g^2 $.
We wish to consider a toroidal compactification to
five dimensions with an $S^1$ of length $2\pi R$ and
a $T^4$ of four-volume $(2\pi)^4 V$.\foot{With
these conventions, T-duality sends $R$ to $1/R$ or $V$ to $1/V$, and S-duality
sends $g$ to $1/g$.}
We will work with the following   near-extremal solution 
labeled by three charges\foot{
This corresponds to the limit $\alpha, \gamma \gg \sigma $
of the solution in \hms, which is the dilute gas region discussed
in the section 3.3. The exact metric has subleading corrections.} \hms , 
given in terms of the ten-dimensional
variables by
\eqn\dil{ e^{-2\phi } = \(1+   { r_5^2\over r^2 }\)
\(1 + {r_1^2 \over  r^2 }\)^{-1},}
\eqn\htn{H= 2 r_5^2 \e_3 + 2 r_1^2  e^{-2 \phi}  *_6 \e_3,}
\eqn\metric{\eqalign{
ds^2 = &
 \( 1 + { r_1^2 \over r^2}\)^{-1/2} \( 1 + { r_5^2 \over r^2}\)^{-1/2}
 \[ - dt^2 +dx_5^2
\right.
\cr
+&
\left.
{r_0^2  \over r^2} (\cosh\sigma dt + \sinh \sigma  dx_5)^2
 +\( 1 + { r_1^2  \over  r^2}\) dx_i dx^i \] \cr
 +& \( 1 + { r_1^2  \over r^2}\)^{1/2}\( 
1 + { r_5^2 \over r^2}\)^{1/2} \left[
\left( 1- {r_0^2 \over r^2 } \right)^{-1} dr^2 + r^2 d \Omega_3^2 \right]
,}}
where $*_6$ is the Hodge dual in the six dimensions $x^0,..,x^5$
and $\e_3$ here is the volume element on the unit three-sphere.
$x^5$ is periodically identified with period $2\pi R$,
$x^i$,  $i = 6,...,9$, are each identified with period $2\pi V^{1/4}$.
The three charges are
\eqn\charges{
\eqalign{
   Q_1 &= {V\over 4\pi^2 g\apm}\int e^{2 \phi} *_6H , \cr
   Q_5 &= {1\over 4\pi^2 g\apm} \int H ,
\cr
  n &= {RP} ,
}}
where $P$ is the total momentum around the $S^1$.
All charges are normalized to be integers and taken to be positive. 
In terms of these charges
the parameters of the solution read
\eqn\paramsol{
r_1^2 = {g Q_1 \apm \over V }, ~~~~~~
r_5^2 = {g Q_5 \apm },~~~~~~r_0^2 {\sinh 2  \sigma \over 2}
= { g^2 n \apm \over R^2 V },~~~~~~
r_n^2 \equiv r_0^2 \sinh^2  \sigma, 
}
and we are in the dilute gas region defined by
\eqn\smallp{
r_0 , r_n \ll r_1 , r_5 .
}
The extremal limit is $r_0 \to 0, ~\sigma \to \infty$ with 
$n$ held fixed. 

The entropy and  energy are
\eqn\extl{\eqalign{E &
= { \pi \over 
4 G_5 } [ r_1^2 + r_5 ^2 + {r_0^2\cosh 2\sigma \over 2 } ]
%= {1 \over g^2}\left[  {R gQ_1}  +
%{  R V gQ_5}\right] + { 2 \pi r_0^2\cosh 2\sigma \over G_5 
= {1 \over g^2}\left[  {R gQ_1}  +
{  R V gQ_5}+{g^2n \over R} +{VRr_0^2e^{-2\s}\over 2}\right]   ~,\cr
S&= { A \over 4 G_5 } = 
{ 2 \pi^2 r_1 r_5 r_0 \cosh \sigma \over  4 G_5 }~~,}}
where the five-dimensional Newton's constant is $G_5=g^2\pi /4VR$.

The D-brane representation of this state involves a bound state 
of $Q_5$ fivebranes wrapping $T^4\times S^1$ and $Q_1$ 
onebranes wrapping the $S^1$. The excitations of this bound state
are approximately 
described by transverse oscillations (generated by open strings
attached to the D-brane),
within the fivebrane,
of a single effective string  wrapped $Q_1Q_5$ times \mss\ around the 
$S^1$. These oscillations  carry the momentum $n$ and are described by
a gas of left and right movers on the string. Equating the energy of
this
gas to $ {n \over R }+{RVr_0^2e^{-2\s } \over 2g^2}$ and its 
momentum to ${n\over R}$ we can determine the total energy 
carried by the right and the left movers. Their entropies match
\extl\ in the dilute gas region \smallp\  \ghas. The left and
right 
moving oscillations are governed by  effective left and right moving 
temperatures  
\eqn\tleft{
T_L = {1\over \pi } {r_0 e^\sigma  \over 2 r_1 r_5 } ,~~~~~~~~~~~
T_R = {1\over \pi } {r_0 e^{-\sigma}  \over 2 r_1 r_5 }. ~~~~~~
}
Notice that in the dilute gas region 
$T_L,T_R \ll  1/r_1 , 1/r_5 $.
%In we take $r_1 \sim r_5 $ but the result is the
%same 
%if they are different (but still in the region \smallp )
%It is natural to say that the gravitational size of the black hole is 
%of the order of $r_1$.
%We also take $V \sim  $.  

\newsec{The Classical Limit and Expansion Parameters}
 
We consider a number of different expansions in this paper. The semiclassical 
expansion is a quantum expansion about a classical limit in which black hole 
radiation is suppressed. Large (relative to the string length) 
black holes can be analyzed in sigma model perturbation theory, while small 
black holes can,  in favorable cases, be analyzed in 
D-brane perturbation theory. 
Those favorable cases are when the parameters are in 
the dilute gas region. 
Both large and small black holes have classical limits (and 
as explained in \bdpss\ and in section 3.2 
below, both deserve the name black hole). In this 
section we describe these 
regions and expansions in detail.

\subsec{The Classical Limit}

In the classical limit the action becomes very large so that 
the stationary phase approximation can be applied. Since the
action \fds\ has an explicit $1/g^2$ prefactor, the limit 
$g\to 0$ with the fields held fixed is a classical limit.
Noting the explicit factors of $1/g$ in the definitions 
\charges\ of the integer charges, as well as the explicit 
$1/g^2$ in the definitions of the energy $E$ and momentum $P$, 
this is equivalent to 
\eqn\clim{\eqalign{g &\to 0,
\cr {\rm with} &~~gQ_1, ~~gQ_5, ~~g^2 n ~~{\rm fixed.}}}
Hence the quantization conditions on 
integer charges imply that they diverge in the classical limit,
as expected. Noting the relations \paramsol, one may 
equivalently define the classical limit with $r_1, r_5$ and $r_n$ held fixed. 

The classical solutions depend only on the products $gQ_1, ~~gQ_5$
and $ ~~g^2 n$ and so are  finite in the limit \clim.
The standard definitions of the ADM energy and momentum 
involve explicit 
$1/g^2$ factors and so diverge. This divergence can be 
eliminated by a change of units
accompanying the limit.
However, the entropy diverges like $1/g^2$, and 
is a dimensionless number which cannot be rescaled.

\subsec{Large and Small Black Holes}
It follows from the metric \metric\ that $gQ_1, ~~gQ_5$
and $ ~~g^2 n$ are the characteristic (squared) sizes of the  
black hole. Hence the black hole is large or small 
depending on whether these quantities are large or small
relative to the string scale. One might question the use of the phrase
black hole to refer to something smaller than the string scale. 
This name is appropriate because the black holes are black 
independently of their size. Because of the divergence 
in the classical limit of the entropy \extl, it costs 
an infinite amount of entropy for the black hole to lose any finite
fraction of its mass in outgoing radiation \hawkirr\ \tata. 
Hence the second law prohibits radiation
from escaping, and black holes are black in the classical limit \clim\
independently of their size.    

Closed string perturbation theory naturally treats the fields 
$\phi,~~g_{\mu\nu}$ and $H$ as order
one. Hence, noting the explicit factors of $1/g$ in \charges,  
it is an expansion in $g^2$ with $gQ_1, ~~gQ_5$
and $g^2 n$ fixed. The classical limit \clim\ is therefore described 
by genus zero closed string theory.  A primary tool for 
analyzing black hole solutions in classical closed string theory 
is the $\alpha^\prime$ expansion. The solutions \dil -\metric\ are solutions of
the leading order $\apm$ equations. They are characterized by 
the squared length scales $gQ_1\apm , ~~gQ_5\apm $
and  $ ~~g^2 n\apm $.  The $\alpha^\prime$ expansion 
is valid when these are large 
in string units:
\eqn\cspert{gQ_1>1, ~~gQ_5>1,~~g^2n>1.}

D-brane perturbation theory 
on the other hand involves both open and closed string
loops. Closed string loops have factors of $g^2$, while open string 
loops have factors of $gQ_1$ or $gQ_5$, corresponding to the fact
that the open string loops can end on any of the D-branes. Hence the 
classical limit \clim\ is a large $N$ limit 
of the open string field theory. Closed string loops are 
suppressed.
The large $N$ limit is the sum over planar open string diagrams with 
holes in them. In practice this series cannot be summed. A primary tool
for analyzing the large $N$ limit is open string perturbation theory.
This is good if 
\eqn\dpert{gQ_1<1, ~~gQ_5<1, ~~g^2n<1.}
The last condition arises because, at the price of a power of 
$g^2$, a Feynman diagram can pick up a power of $n$ by hooking propagators to 
the momentum in the external state \bdpss.  

Hence the classical limit \clim\ may be characterized either 
by the classical genus zero closed string theory or by the large 
$N$ limit of the quantum D-brane open string theory. In 
general factorization of large $N$ matrix elements implies that 
every large $N$ theory is describable by a classical master field.
In the present context this classical master field is provided by the
closed string theory. These two different representations of the 
limit \clim\ are useful in different regimes of the couplings
according to \cspert\ and \dpert. The closed string theory is good 
for large black holes (relative to the string scale) while the 
D-brane field theory is good for small black holes. This relation is
being explored in \bdpss. 

In summary the limit \clim\ defines a semiclassical limit for both 
small and large black holes. The semiclassical Hawking calculation is 
well justified in the large black hole region \cspert. D-brane 
perturbation theory is well-justified in the region \dpert.

\subsec{The Dilute Gas Region}

A further condition is needed in order to simplify the calculation of 
non-extremal entropies and decay rates in the string picture. 
In general the left and right moving oscillations on the string interact, 
and their entropy and energy spectrum 
is not exactly that of a free two-dimensional gas.
We can understand heuristically when this free gas approximation will 
break down as follows (a more precise discussion can be found in \hms).
Since these left and right movers represent oscillations of the 
string, we see that a necessary condition is
that the typical amplitude $A$ of the oscillations is smaller than the 
typical wavelength $ \lambda $. This is the standard small amplitude
approximation
for propagating waves. 
The total energy in these oscillations is $n/R$ if they are all 
left moving. If this
energy  is carried by an effective string of length $Q_1Q_5 R$ and
 tension $1/Q_5g$ we get the relation
\eqn\amplit{
E = {n \over R } \sim {Q_1 R \over g} 
\left( A\over \lambda \right)^2~. 
}
Demanding that $A \ll \lambda $ we find
\eqn\lsrt{
{ n g \over R^2 Q_1 } \sim { r_n^2 \over r_1^2 } \ll 1~~~~~{\rm or}~~~~~~~
r_n \ll r_1. }
This result does not depend on how the strings are wound, or 
whether they form a long string of length $RQ_1 Q_5 $, although the 
precise momentum quantization condition does \mss . 
A T-dual analysis gives the  
condition $r_n \ll r_5$. Analogous considerations with right-movers gives 
$r_0\ll r_1,r_5$. 

While we will not attempt to do so in this paper, it may be 
possible to drop the 
restriction to the dilute gas region using ideas introduced 
in \cm  . It is possible to 
view the corrections to the entropy away from the dilute gas 
limit as arising from 
antibranes or closed ``fractional'' strings \jmn . 
The form of these corrections is highly constrained 
by duality and it is 
possible  - with some assumptions - to account for the all the 
entropy everywhere in the moduli space in this fashion \hms . 
Possibly this approach could be used to extend the 
results of this paper over the entire moduli space at low energies.

\newsec{Neutral Scalar  Emission}

In this section we will compute the decay rate into neutral scalars 
of an excited black hole using the Hawking 
formula including greybody factors and compare it to the 
corresponding perturbative 
string decay rate. 

The greybody factor in the Hawking formula \hgf\ for the emission rate of a 
given type of outgoing particle at energy $\omega$ equals the 
absorption cross section $\s_{abs}$ for the particle incoming 
at energy $\omega$ \hawkb \bd. 
Greybody factors were computed for the emission of various particles
in \page \unruh , but not for black holes in the dilute gas
approximation.

We first compute this absorption cross section for neutral 
scalars incident on  
the near  extremal black hole \metric .
The calculation is done by solving the Klein Gordon equation
describing 
the propagation of the particle on the fixed black hole background.
The classical wave equation is the laplacian in the five 
dimensional Einstein metric
\eqn\class{
\left[
{ h \over r^3} { d\over d r} h r^3 {d \over d r } + \omega^2 f 
\right] R = 0,
} 
$$ f = ( 1 + { r_n^2 \over r^2 } )( 1 + { r_1^2 \over r^2 } )
( 1 + { r_5^2 \over r^2 } ), 
$$
\eqn\defgis{ h = 1 -{ r_0^2\over r^2 },
}
where $\omega$ is the energy of the wave. 
In this theory there are many scalars. The wave equation \class\
describes
the interaction of a scalar that does not couple to the gauge field
strength. One example of such scalar, studied in \dm,  is 
an off-diagonal component
({\it e.g.} $h_{78}$)
of
the internal metric tangent to the $T^4$.
For the other scalars both the
wave equation \ka\ and the D-brane calculation require modifications. 
The function $f$ is the
product of the three harmonic functions characterizing the black hole
and
$r_0$ is the non-extremality parameter.
We assume that we are in the dilute gas region \smallp, together with
the low energy condition
\eqn\conditions{
  \omega r_5 \ll 1
}
while we treat the ratios $ \omega /T_{R,L},~~r_1/r_5,~r_0/r_n$ 
as order one.

The absorption cross section is usually computed from 
solutions to the wave equation which have unit 
incoming flux from infinity and 
no outgoing  flux from the past
horizon. The absorption cross section is then the difference
of the incoming and outgoing flux at infinity. This difference will 
be small at low energies. Equivalently one may
% impose the boundary
%condition that there is no  outgoing flux on  the past horizon 
%and 
compute the ratio of the ingoing flux at the future horizon to
the incoming flux from past infinity. We shall follow this latter 
approach as it avoids finding the small difference of two 
larger quantities. 

The wave equation \class\ does not appear to be analytically 
soluble. 
The solutions can be approximated by matching near and far zone solutions. 
We divide the space in two regions: the far zone $r > r_m $
and the near
zone $ r <r_m $, where $r_m$ is the point where we will match the
solutions. $r_m$   is chosen so that 
\eqn\condrm{ r_0,r_n \ll r_m \ll r_1,r_5 ,~~~~~~~~~~
\omega r_1 {r_1 \over r_m } \ll 1. 
}
Notice that the last condition is automatically satisfied, given
the others, since $\omega \sim T_{L,R}$.

In the far zone after the change of variables to $\rho = \omega r $,
and $R = r^{-3/2} \psi $ the equation becomes 
\eqn\freeeq{ 
 { d^2 \psi \over d \rho^2 }  + 
\left[ 1 + { -3/4 + \omega^2(r_1^2 + r_5^2 ) \over
\rho^2 } + { r_1^2 r_5^2 \omega^4  \over \rho^4 } 
+ \cdots \right] \psi =0. 
}
For $r > r_m $, we see from  \condrm , \conditions  ~that 
\freeeq\ reduces to 
\eqn\freeequ{
{ d^2 \psi \over d \rho^2 }  + 
( 1 -{  {3}  \over  4\rho^2 } ) \psi
=0.
}
Two independent solutions are the Bessel and Neumann functions 
\eqn\twosol{ 
\eqalign{
F = & \sqrt{ \pi \over 2 } \rho^{1/2} J_1(\rho),\cr
G = & \sqrt{ \pi \over 2 } \rho^{1/2} N_1(\rho).
}}
The solution can be expressed as $R = { 1 \over r^{3/2} } ( \alpha F +
\beta G )$ and has the following asymptotic expansion for very large
$r$,
very far from the black hole
\eqn\veryvery{
R = {1 \over r^{3/2} } \left\{ e^{i\omega r} [
{ \alpha \over 2 }  e^{-i3 \pi/4}  - { \beta \over 2} e^{-i\pi/4} ] 
+ e^{-i\omega r} [ { \alpha \over 2 }e^{ i3 \pi/4} - { \beta \over 2}
 e^{i\pi/4} ] \right\}, 
}
while for small $r$, $r\sim r_m $,  we have have to use the small $\rho$ 
expansion of the Bessel and Neumann functions
\eqn\expanbes{
\eqalign{
J_1(\rho) \sim & { \rho \over 2 }, \cr
N_1(\rho) \sim & { 1 \over \pi} \left[ \rho ( \log \rho + c )
-
 { 2 \over \rho} \right],
}
}
where $c$ is a numerical constant. 
Using \expanbes \twosol ~we get for small $r$ 
\eqn\expansmall{
R = \sqrt{\pi \over 2 } \omega^{3/2} \left[
 { \alpha \over 2 } + { \beta \over \pi } ( c + \log(
\omega r) - { 2 \over \omega^2 r^2 } ) \right].
}
At $r=r_m$ the term multiplying $\beta $ is very large. 
We will see that this will imply that $\beta \ll  \alpha $.
% Notice that our treatment is slightly different than D-M. 

In the near zone we have the equation
\eqn\near{
{ h \over r^3} { d\over d r} h r^3 {d R\over d r } + [{(\omega r_n r_1
r_5)^2 \over r^6 } + { \omega^2 r_1^2 r_5^2 \over r^4 }] R =0,
}
which is valid for  $r <r_m $. 
Defining a new variable $
v = r_0^2/r^2
$
the equation becomes
\eqn\neartwo{
(1-v) {d \over d v} (1-v) {d R \over d v} + (D + {C \over v})R =0,
}
where 
\eqn\definitions{
D = \left( {\omega r_1 r_5 r_n \over 2 r_0^2 } \right)^2,~~~~~~~~~
C = \left( {\omega r_1 r_5  \over 2 r_0 } \right)^2 ~.
}
The horizon is now at $v=1$ and the matching region ($r\sim r_m$)
 is at small $v$. 
Very close to the horizon we can change variables
to $ y = -\log(1-v)$ and the equation becomes
\eqn\nearhor{
{d^2 R \over d y^2 } + (C+D) R =0,
} 
which has the solutions 
\eqn\nearhsol{
R_{in} = e^{ - i \sqrt{C+D} \log (1-v) },~~~~~~~~~
R_{out} =  e^{ + i \sqrt{C+D} \log (1-v) }.
}
$R_{in} ~(R_{out})$ is the ingoing (outgoing) solution 
at the horizon. 
The boundary condition is that for $v\sim 1$ the solution should
behave like 
\eqn\bound{
R = A e^{-i\sqrt{C+D}  \log(1-v) },
}
where $A$ is a constant to be determined later.
Now let us solve the equation \neartwo . We define  new variables $z$
and
$F$ by 
\eqn\newvar{
z = (1-v),~~~~~~~~~~~~ R = A z^{-i(a +b)/2} F,
}
where $a,b$ will be fixed below to simplify the equation.
Substituting \newvar\ into the equation \neartwo\ 
we obtain a hypergeometric equation for $F$
\eqn\hyper{
z(1-z) {d^2 F \over d z^2 } + [\gamma - (1 -ia - ib )z] 
{d F \over d z} + ab F =0,
}
where $\gamma =(1 -ia - ib )$ and 
 $a,b$ are defined by the equations $ (a+b)^2 = 4 (C+D)$ and $ab
= C$. This yields 
\eqn\defab{\eqalign{
a =& {\omega r_1 r_5 e^{\sigma} \over 2 r_0 } = {\omega \over 4 \pi
T_R },\cr
b = & {\omega r_1 r_5 e^{-\sigma} \over 2 r_0 } = {\omega \over 4 \pi
T_L },
}}
where we have used \tleft . 

Equation \hyper\ has a one parameter family of normalized solutions.
Imposing the boundary condition \bound ~and using the definitions \newvar ~
we find that the desired solution is 
\eqn\solution{
R = A z^{-i(a +b)/2} F(-ia,-ib,1- ia - ib  -\epsilon, z),
}
where $\epsilon$ is a regularization parameter we introduce for
later convenience. Note that $F(\alpha, \beta,\gamma,
0)=1$ while the other solution to \hyper\ behaves as $z^{1-\gamma} F(...) =
 z^{i(a+b)} $ corresponding to an outgoing wave.
To determine the form of the solution for small $v$ we express the
$F$ in terms of $1-z=v$ using the hypergeometric relation
\eqn\transf{\eqalign{
F(-ia,&-ib, 1 -ia - ib  -\epsilon, z) = 
{ \Gamma(1 -ia -ib -\epsilon) \Gamma(1-\epsilon) \over
\Gamma(1 - ib -\epsilon ) \Gamma(1 - i a - \epsilon ) }
F( -ia, -ib, \epsilon, v) \cr
  +&  v^{1-\epsilon} 
{\Gamma( 1  -ia -ib -\epsilon) \Gamma(-1+\epsilon) \over
\Gamma( -ib ) \Gamma( -i a  ) } F(1- ia -\epsilon, 1-ib -\epsilon,
2-\epsilon, v )
.}}
Note that the singularities cancel for $\epsilon \rightarrow 0$.
The resulting expression has  the following expansion for 
small $v$
\eqn\result{
F \sim E +  v( G  + G'  \log v  ) +...,}
where the constants $E,~G,~G'$ are independent of $v$ but depend on
$a$ and $b$. 
The contribution to the $v$ independent  term comes only 
from the first term in
\transf ~:
\eqn\edefnition{
E = { \Gamma(1 -ia -ib )  \over
\Gamma(1 - ib ) \Gamma(1 - i a ) }.
}
Now we match the solutions \result ~and \expansmall ~together with 
their first derivatives at $r =r_m$.
We obtain the equations
\eqn\match{\eqalign{
\sqrt{ \pi \over 2 }  \omega^{3/2} \left[
 { \alpha \over 2} + { \beta \over \pi } ( c + \log(
\omega r_m) - { 2 \over \omega^2 r^2_m } ) \right]
= &A \left[ E +  v_m( G  + G'  \log v_m ) \right], \cr
\sqrt{ \pi \over 2 }  \omega^{3/2}
{ \beta \over \pi } ( 1 + { 4  \over \omega^2 r^2_m } ) =&
-2 A  {v_m} ( G + G' \log v_m + G' ).
}}
Using \condrm\ and 
$v_m = {r_0^2\over r^2_m}$ ~we 
conclude that $\beta/\alpha \ll 1 $.
We can also neglect the term involving $\beta$ in the first equation in
\match .~ We then obtain
\eqn\defalpha{
\sqrt{\pi \over 2} \omega^{3/2} {\alpha \over 2} = A
E,~~~~~~~~~~~{\beta\over \alpha }
\ll 1 ,
}
% The fact that $G v_m \ll 1 $ implied that $\beta/\alpha \ll 1 $
so that we do not need $\beta$ to compute the incoming flux.
Notice that we are basically matching the free particle solution 
$\beta =0$ to the amplitude of the solution inside the throat. 
This is reasonable considering that the wavelength is much larger
than the size of the black hole. 

The conserved flux is given by  
\eqn\flux{
f = {1 \over 2i } [ R^* h  r^3 { d R \over d r} - c.c. ].
}
The incoming flux from infinity, as 
calculated from \flux , \veryvery ,\defalpha, is 
\eqn\incomingf{
f_{in} = - { \omega | {\alpha \over 2 }|^2 }.
}
The flux into the black hole at the future horizon is 
\eqn\absobed{
f_{abs} = {1 \over 2i } [ R^* 2 r_0^2 (1-v)  { d R \over d v} - c.c. ]
= - r_0^2 {( a+b) } |A|^2.
}
The absorption cross section for the S-wave is then \defalpha 
\eqn\sigmaswave{
\sigma^S_{abs} = { f_{abs} \over f_{in} } =  r_0^2 { (a + b ) \over \omega  }
|E|^{-2} \omega^3 { \pi \over 2 }. 
}
The absorption cross section for a plane wave of 
frequency $\omega$ is related to the S-wave
cross
section by (see \dm ~(6.29 -6.31 ) )
\eqn\absor{
\sigma_{abs} = { 4 \pi \over \omega^3 } \sigma^S_{abs} = 
2 \pi^2 r_0^2 
{ (a + b ) \over \omega }|E|^{-2}.
}
Next we compute $|E|^2$. Using the identity 
\eqn\modgamma{
|\Gamma( 1 - ia ) |^2 = { \pi a \over \sinh \pi a },
}
we find 
\eqn\modeis{
{ 1 \over |E|^2 } =  2 \pi { ab \over (a + b)  } { (e^{2\pi (a+b)} -1 )
\over (e^{2 \pi a } -1 ) (e^{2 \pi b } -1 ) }.
}
Inserting the values of $a,b$ from \defab\ ~in \modeis\ and then in
\absor , 
%remembering that the area of the horizon is $A_H = 2 \pi^2 
%$r_1r_5 r_0 \cosh \sigma $ 
we obtain the final expression for the
absorption
cross section
\eqn\absorfinal{
\sigma_{abs} = 
2 \pi^2 r_1^2 r_5^2 {\pi \omega \over 2 } 
{
e^{ { \omega \over T_H }}  -1 \over (
e^{{ \omega \over 2 T_L } } -1 )( e^{ { \omega \over 2 T_R } } -1 )}
,}
where the Hawking temperature is
\eqn\thaw{
{ 1 \over T_H } = {1\over 2 } (  { 1 \over T_L} +  {1 \over T_R} ).
}
According to Hawking \bd\ the emission rate is equal to
\eqn\hawclass{
\Gamma_H = \sigma_{abs} { 1 \over e^{ { \omega \over T_H } }  -1 }
\dfourk = 2 \pi^2 r_1^2 r_5^2  {\pi \omega \over 2 } { 1 \over (
e^{ { \omega \over 2 T_L } } -1 ) } { 1 \over (
e^{ { \omega \over 2 T_R } } -1 ) } \dfourk
}

The D-brane emission rate in the dilute gas region 
is given by \dm\
\eqn\dbranehaw{
\Gamma_{D} = 2 \pi^2 r_1^2 r_5^2  {\pi \omega \over 2 } 
\rho({ \omega \over 2 T_L })\rho({ \omega \over 2 T_R }) \dfourk =
2 \pi^2 r_1^2 r_5^2  {\pi \omega \over 2 } { 1 \over (
e^{ { \omega \over 2 T_L } } -1 ) } { 1 \over (
e^{ { \omega \over 2 T_R } } -1 ) }\dfourk .
}
The factors of $\rho_{L,R}$ come from the thermal ocupation factors.
We see that this expression agrees precisely with \hawclass . 

To recover the results of \dm, we make the further 
approximation $T_R\ll T_L$. One then has  
$T_H=2T_R$, $\omega \sim T_H$, $\rho(\omega/2T_R) \sim
\rho(\omega/T_H)$ 
and $\rho(\omega/2T_L) \sim
2T_L/\omega$.  Using the expression \extl\ for the area, 
the decay rate \dbranehaw\ then reduces to
\eqn\dmd{\Gamma_H =\Gamma_{D}= A_H \rho({\omega \over
T_H})\dfourk .}

\newsec{ Charged Scalar Emission}

Now we turn to the problem of calculating the emission rates
for scalars that carry Kaluza-Klein charge.\foot{The $T_R \ll T_L$ limit of 
the results of this section were obtained in \gk. } 
In five dimensions they are massive  
particles with mass saturating a BPS bound.
In six dimensions they are massless particles with momentum along 
the direction of the string. Hence in the limit of large 
$R$ the problems of neutral and charged emission are 
related by a boost along the direction of the string. 
Since both the string and the spacetime picture are 
boost invariant in this limit, we expect the agreement found in 
the neutral case to extend to the charged case. 

We begin by calculating the emission rate in the string picture. 
The string calculation is a simple extension of the calculation
in \dm\ in which one relaxes the condition that the 
interacting pair of left and right moving oscillations  have opposite
momenta. The emission rate in the dilute gas region is 
\eqn\crossd{\eqalign{
\Gamma_D =
{ d^4 k \over (2 \pi )^4 }
 { 8 \pi^3 r_1^2 r_5^2\over k_0 } \int_0^\infty { d p_5 \over 2 \pi p_0 }
\int_0^\infty
 { d q_5 \over 2 \pi q_0 } ( 2 \pi )^2 &\delta( k_0 -p_0-q_0) \delta(
k_5 - p_5 + q_5 )\cr & ( p.q/2 )^2 \rho( p_0/T_L ) \rho(q_0/T_R),}
}
where  $(k_0,k_5,\vec k) $ is the momentum of the incoming particle
and $(p_0,p_5),(q_0,-q_5)$ are the momenta of the left and right movers on
the string. 
$k_5$ is the charge from the five-dimensional point of view and 
is of the form $m/R$ for some integer $m$. 
Since they are massless particles $p_0=p_5,~q_0=q_5$.
Momentum conservation implies that $p_0 = (k_0+k_5)/2$, $q_0 =
(k_0 - k_5 )/2 $. 
Evaluating the integrals in
\crossd\ we find
\eqn\crossdfin{
\Gamma_D  =  2 \pi^2 r_1^2 r_5^2 {\pi ({k_0}^2 -{k_5}^2 ) \over 2 k_0 } 
 { 1 \over (
e^{ { k_0 + k_5 \over 2 T_L } } -1 ) } { 1 \over (
e^{ { k_0 - k_5 \over 2 T_R } } -1 ) } \dfourk.
}
Note that we do  NOT assume that
 $p_0 \ll  T_L $.

Now we turn to the Hawking calculation. 
 We first   calculate the  absorption cross section by solving  the 
Klein Gordon wave equation on this background. 
It is easier to think of the background as six-dimensional. 
The six-dimensional dilaton $Ve^{-2\phi}$ is  constant  \hms , so that the
six-dimensional Einstein and string metrics are equivalent.
For low energies
the dominant contribution to the cross section  comes from the S-wave, so
that  the
Klein Gordon equation becomes
\eqn\kg{
\left(
G^{00} \partial_0^2 + 2 G^{05} \p_0 \p_5 + G^{55} \p_5^2 \right) \Phi
+ { 1 \over \sqrt{G} } \p_r ( \sqrt{G} G^{rr} \p_r \Phi ) =0, 
}
with the near-extremal metric of \hms . We work in the dilute gas region
$r_0, r_n \ll r_1,r_5 $. 

Defining $\Phi = e^{-ik_0 t  - i k_5 x^5 } R(r) $ we obtain the
radial equation  
\eqn\kgr{
(1 + { r_1^2 \over r^2} )(1 + { r_5^2 \over r^2} )\left[
{k_0}^2 - {k_5}^2 + (k_0 \sinh \sigma   - k_5 \cosh \sigma   )^2 
{ r_0^2 \over r^2  } \right] R
+ { h \over r^3 } {d\over d r} ( h r^3 {d R \over dr } ) =0,
}
where $h$ is defined in \defgis . We define  new variables
\eqn\newpar{
\omega'^2 =
 {k_0}^2 - {k_5}^2 , ~~~~~~~~~e^{\pm \sigma'} = 
e^{\pm \sigma}{(k_0 \mp k_5) \over \omega'} ,~~~~~~~~~~~~
r'_n = r_0 \sinh \sigma'.
}
Reexpressing \kgr ~ in terms of these new variables, we find it 
reduces to the equation \class\ governing neutral absorption 
with the substitutions $\omega\to \omega'$ and 
$r_n \to r'_n$. 
%Notice that even if we
%had originally $r_n \gg r_0$ we could have $r'_n \sim r_0$ due to
%the parameter redefinition \newpar . 
Notice that the parameters
$r_0, r_1, r_5 $ are unchanged.
Hence the results of the previous section \absorfinal\ imply 
that the absorption cross section is
\eqn\crossclass{
\sigma_{abs} =   
2 \pi^2 r_1^2  r_5^2 { \pi \omega' \over 2} { e^{  \omega'/T'_H} -1
\over ( e^{ {\omega'\over 2T'_L} } -1 )
( e^{{\omega'\over 2T'_R} } -1 ) }.
}
Rewriting this in term of the original variables
\eqn\temper{\eqalign{
{ \omega' \over T'_L } =& { \omega' \over T_L }e^{\sigma -\sigma'} =
{ k_0 + k_5 \over T_L } ,\cr
{\omega' \over T'_R } = & { k_0- k_5 \over T_R } ,\cr
{ \omega' \over T'_H } = & {k_0 + k_5 \over 2T_L } + {k_0 -k_5 \over
2T_R } = { k_0 -\phi k_5  \over T_H }, 
}}
where $\phi = \tanh \sigma $ is the electrostatic potential at the 
horizon, $\phi= A_0(r_0), $ with $~~A_0(r)
 = { r_0^2 \sinh 2 \sigma \over 2 r^2}
( 1 + { r_0^2 \sinh^2 \sigma \over r^2 } )^{-1} $, 
we finally obtain for the classical absorption cross section
\eqn\absorfin{
\sigma_{abs} = 2 \pi^2 r_1^2 r_5^2 { \pi \sqrt{ {k_0}^2 - {k_5}^2 } \over
2}
{ e^{ k_0 - k_5 \phi\over T_H  } -1 \over (e^{k_0 + k_5 \over 2 T_L } -1 ) 
(e^{ k_0 - k_5 \over 2 T_R } - 1 ) }.
}
The Hawking rate for charged particles is in general\thooft\
\eqn\hawcha{
\Gamma = v \sigma_{abs} { 1 \over e^{ k_0 - k_5 
\phi \over T_H } -1
}\dfourk ,
}
where the factor of the particle velocity, $v = \omega'/k_0$,
is a kinematical factor and
$\phi$ is the scalar potential at the horizon. 
Inserting \absorfin ~in \hawcha ~we obtain 
\eqn\hawfincha{
\Gamma = 2 \pi^2 r_1^2 r_5^2 { \pi ({k_0}^2 -{k_5}^2 ) \over 2 k_0 }
{ 1 \over (e^{k_0+k_5 \over 2 T_L } -1 )}
{1 \over (e^{k_0 -k_5 \over 2 T_R } -1 )}\dfourk ,
}
which agrees precisely with the string result \crossdfin .

\newsec{Scalar Absorption }

In the preceding two sections we calculated and compared emission rates
in the string and Hawking pictures. It is also of some interest 
to consider absorption rates, which have the qualitative difference
that they do not vanish in the classical limit\foot{As discussed in 
\hawkirr\ \tata\ this apparent time irreversibility follows 
from the entropy formula and the second law.}. 

Pieces of the calculation already appeared in the preceding sections
and it is not hard to see the agreement 
directly. An illuminating subtlety 
is that the thermal factors $\rho_{L,R}$ appearing in the emission 
rate are replaced by 
$\rho_{L,R} +1 $ in the absorption rate, 
corresponding to the matrix element of a 
bosonic creation rather than annihilation operator.
The classical absorption cross section as computed from the 
classical wave equation is equal to the 
string absorption cross section minus the emission
rate for that mode. This difference is
just proportional to $ (\rho_L +1 )(\rho_R+1 ) -\rho_L \rho_R =
\rho_L\rho_R/\rho_H$. This is precisely the combination of thermal 
factors we see appearing in 
the classical calculations done above \absorfinal\ \absorfin .
Hence the appearance of this particular combination of factors is already
necessary for agreement in the classical limit.  

It is also interesting to consider the absorption cross section in the case 
$T_R=T_H=0$, which corresponds to absorption by an 
extremal black hole. One finds 
\eqn\sigmaabs{
\sigma_{abs} = A_H { \omega \over 2 T_L } { 1 \over ( 1 - e^{- \omega /2
T_L } )}.
}
Notice the appearance of the thermal factor $\rho(\omega/2T_L) +1 $
which has no simple explanation from the spacetime black hole
picture, but
is obvious from the string perspective. This is a salient example of
how the classical greybody factors ``know'' about the string.

\newsec{Evolution of a Near-Extremal Black Hole}

In this section we compare the rates of charge 
and neutral emission, and discuss the problem of measuring 
the quantum state of a black hole with scattering experiments. 

We first consider the decay rate due to charged emission.
A near-extremal black hole with excess energy $\dE=Vr_0^2e^{-2\s}/2g^2$ 
above extremality 
has a Hawking temperature 
\eqn\thw{T_H= {1\over \pi } \sqrt{2 \dE \over Q_1Q_5R}.}
For small $\dE$,
$T_H$ is smaller than the mass $1/R$ of the lightest charged 
state and  $2 T_R\sim T_H$. Hence the 
outgoing charged particles are all highly 
non-relativistic. Their kinetic energies are approximately
$k_0-k_5\sim {\vec k^2\over 2k_5 }$.
It then follows from the thermal factors in 
\hawfincha\ that the kinetic energies are of
order $T_H$ (rather 
than the total energies as in the neutral case). 
Emission of a charged particle decreases both the total
energy and the charge of the black hole. The excess energy $\dE$ 
is decreased only by the kinetic energy of the outgoing particle 
which is just $T_H$.

With these approximations we can calculate the rate of decrease of $\dE$
due to emission of particles with charge $k_5$ from \hawfincha\
\eqn\energyloss{
{d \dE \over d t } = 
\int    
{\vec k^2\over 2k_5 }  \Gamma   = {\pi^2 \over 60} A_H T_H^4 { {
k_5}^2 \over T_L } { 1 \over e^{k_5/T_L} -1 } ~ .
}
Note that typically $k_5 \sim T_L $, where $Rk_5$ is an integer, 
when $RT_L={1\over  \pi}\sqrt{n \over Q_1Q_5}$ is greater than one. 
This rate is exponentially suppressed by the factor $e^{- k_5/T_L}$ for
$k_5 \gg T_L $.
This exponential supression is due to the fact 
that the emission of a particle with charge $k_5$
reduces the entropy of the extremal black hole by $\Delta S = k_5/T_L
$, and so must be accordingly suppressed. For large $RT_L$ the total 
emission rate for all charges can be approximated by an integral 
of \energyloss\ over positive $k_5$ :
\eqn\ergyloss{{d \dE \over d t } \sim
 {\pi^2\zeta(3) \over 30} A_H T_H^4 T_L^2 R ,~~~RT_L\gg 1~.}
For small $RT_L$ charge emission is dominated by the minimal value 
$k_5 = 1/R$ 
\eqn\eness{
{d \dE \over d t } \sim 
{\pi^2 A_H T_H^4 \over 60R^2 T_L } e^{-1/RT_L} ,~~~RT_L\ll 1 ~ .
}

For neutral emission the integrals yield \dm
\eqn\zerocharge{
{d \dE \over d t } = {3 \zeta(5) \over  \pi^2 } A_H T^5_H~.
}
This expression has one more power of $T_H$ in it than the one 
for the 
charged emission. Hence at sufficiently low energies 
charge emission always dominates. This is 
because there is more phase space available to the  the 
massive charged particles. However, for 
small $RT_L$ charged emission is exponentially suppressed 
and the energies at which it dominates over 
neutral emission become exponentially small. Hence 
charge emission dominates in some regimes while neutral emission
dominates in others. 
 
Next let us consider the rate of charge loss by the black hole 
in the region $RT_L \gg 1$ where charge emission dominates.
Since the black hole decays by emitting charged particles that
carry charge of the order of $k_5 \sim T_L $ and kinetic 
energy $ \delta\dE
\sim T_H $ we conclude that in a typical emission process
\eqn\chargeenerg{
{\delta n \over \delta \dE} \sim { R k_5 \over \delta \dE } \sim { R  T_L \over
T_H } \sim\sqrt  {nR \over \dE .}
}
Integrating this equation we find that by the time $\dE$ decays to
zero 
\eqn\deltan{
{\Delta n \over n } \sim { \Delta S \over S, }
} 
where  $ \Delta S $ is the entropy carried away by 
the charged Hawking radiation.

Now let us consider in this light the problem of measuring 
the quantum microstate of a black hole. 
%Suppose that we had an extremal black hole whose microstate was initially
%unknown, so it is in a mixed state with corresponding (fine-grained) 
%entropy $S_{BH}=2\pi\sqrt{Q_1Q_5n}$.
We might try to measure the microstate by exciting 
it (perhaps repeatedly) with low energy quanta 
and measuring the outgoing charged radiation resulting from the decay.
According to Hawking the outgoing radiation 
carries no information about the microstate
which cannot be measured. Repeated  experiments 
only produce an ever-increasing amount of 
entropy in the radiation. In the string picture 
there is also some entropy in the outgoing radiation, because it is
entangled\foot{{\it i.e.} the complete quantum state is a 
sum of (rather than a single) products of 
black hole states with states of the radiation.} 
with the quantum state of the black hole (which 
we do not directly measure). However, this 
entanglement entropy can never exceed $S_{BH}$,
where $S_{BH}$ is the logarithm of the number of 
possible black hole states.  
This follows from the triangle 
inequality for fine-grained entropies \arl : $S_A + S_B \geq S_{AB}
\geq |S_A - S_B| $.
In the string picture the entropy in the radiation 
will grow initially but then will saturate 
at a value $S_{max}$ 
which is at most $S_{BH}$. For sufficiently rich 
interactions between the radiation and black hole 
microstates it should be possible to arrange so that 
$S_{max}=S_{BH}$. Since the whole system is unitary 
when 
this saturation occurs the black hole 
microstate is fully correlated with the radiation and  
has effectively been measured. So in order to 
measure the microstate of the black hole - 
and to discern the difference 
between the non-unitary Hawking amplitudes and 
the unitary string amplitudes - 
there must be at least of order $e^{S_{BH}}$  quantum states 
accessible to the radiation so that they 
can carry an amount of information of order $S_{BH}$. This requires a 
large number of experiments. 

As noted above, in the region $RT_L \gg 1$ these extremal black holes 
tend to discharge Kaluza Klein charge 
when they interact. 
Indeed there is a simple relation between the entropy produced and the 
charge lost. We see from \deltan\ that by the time the 
outgoing radiation has enough accesible states to 
determine the quantum microstate of the black hole it has lost all of its 
Kaluza Klein charge. 

On the other hand for $RT_L \ll 1$, one could excite the black hole by
an energy $\dE \gg n/R$ above extremality and still remain within the 
near-extremal and dilute gas regions. In this region, charge emission
is exponentially 
suppressed. According to Hawking, the entropy of the outgoing
radiation
will be of order $\sqrt{Q_1Q_5R\dE}$ which is much greater than 
the original entropy $S_{BH}$ of the black hole. In the string
picture the entropy of the outgoing
radiation can not exceed $S_{BH}$. So this presents a sharp puzzle.

{\bf Acknowledgements}

J. M. wishes to thank C. Callan for discussions in the  early
stages
of this work. 
We also benefited from discussions with S. Das,  G. Mandal,
S. Mathur and L. Susskind. This work was supported in part by 
DOE grants DOE91ER-40618 and DE-FG02-96ER40559.

\listrefs

\bye